\documentclass[aps,prd,showpacs,nofootinbib,floats,floatfix,preprintnumbers,groupedaddress,twocolumn]{revtex4}
\usepackage{bm}
\usepackage{latexsym}
\usepackage{dcolumn}
\usepackage{amsmath,amsfonts,amssymb}
\usepackage{graphicx,epsfig}
\usepackage{amsmath}
\usepackage{fancyhdr}
\usepackage{hyperref}
\usepackage{graphicx,epstopdf}

\hypersetup{ colorlinks   = true, urlcolor     = blue, linkcolor =
blue, citecolor   = blue }

\begin{document}

\title{Universality class of alternative phase space and Van der Waals
criticality}
\author{Amin Dehyadegari$^{a}$\thanks{\color{blue} amindehyadegari@gmail.com}}
\author{Bibhas Ranjan Majhi$^{c}$\thanks{\color{blue}
bibhas.majhi@iitg.ac.in}}
\author{Ahmad Sheykhi$^{a,b}$\thanks{\color{blue} asheykhi@shirazu.ac.ir}}
\author{Afshin Montakhab$^{a}$\thanks{\color{blue} montakhab@shirazu.ac.ir}}

\affiliation{$^a$ Physics Department and Biruni Observatory, College of Sciences, Shiraz
University, Shiraz 71454, Iran\\
$^b$ Research Institute for Astronomy and Astrophysics of Maragha (RIAAM),
P.O. Box 55134-441, Maragha, Iran\\
$^c$ Department of Physics, Indian Institute of Technology Guwahati,
Guwahati 781039, Assam, India}
\date{\today }

\begin{abstract}
A new perspective toward thermodynamic phase space of Reisser-Nordstrom (RN)
black holes in an anti-de-Sitter (AdS) spaces was recently proposed \cite%
{AAA}, where the square of the electric charge $(Q^2)$ of black hole was
regarded as a thermodynamic variable and the cosmological constant
(pressure) as a fixed quantity. In this paper, we address the universality
class and critical properties of any AdS black hole in this alternative
phase space. We disclose the critical behavior of AdS black hole in the
alternative phase space in which a continuous phase transition happens and
in a very general framework, independent of the spacetime metric. Based on
the expansion of the equation of state and Landau thermodynamic potential in
the neighborhood of a critical point in the alternative phase space, we
confirm that the set of values for critical exponents for generic black hole
is analogous to the Van der Waals fluid system. Finally, we reveal that the
scalar curvature in geometry thermodynamic diverges at the critical point of
black hole. Our study shows that the approach here is powerful enough to
investigate the critical behavior of any black holes and further supports
the viability of the alternative viewpoint toward phase space of black holes
suggested in \cite{AAA}.
\end{abstract}

\maketitle


\section{Introduction}

Thermodynamic phase transitions are one of the most intriguing phenomena in
black hole physics which can provide insight into underlying structure of
spacetime geometry. In particular, phase transition of anti-de-Sitter (AdS)
black holes has received much attention since the discovery of the
correspondence between gravity in an AdS spacetime and the conformal field
theory (CFT) living on its boundary. The pioneering work in this regard was
performed by Hawking and Page \cite{HP}, who demonstrated the existence of a
first-order phase transition in the Schwarzschild AdS black hole. According
to the AdS/CFT duality, this phase transition can be interpreted as a
confinement-deconfinement transition in the quark gluon plasma \cite{Witten}%
. Later, the analogy between the small-large black hole phase transition and
the liquid-gas Van der Waals phase transition was reported in \cite{VDW1},
where RN black hole was considered in asymptotic AdS spacetime. Later on, it
was revealed that this similarity happens in the extended phase space of the
RN-AdS black hole in which the cosmological constant ($\Lambda $) is
regarded as a thermodynamic variable corresponding to the thermodynamic
pressure with a black hole's volume as a conjugate quantity \cite{Dolan,PV}.
In the extended phase space, the cosmological constant as a dynamical
quantity can take on arbitrary values in the first law of black hole
thermodynamics where the mass of AdS black hole is interpreted as the
enthalpy \cite{enthalpy}. In recent years, various studies on black holes
phase transition in an extended phase space have been carried out such as
zeroth-order phase transition \cite{NAAA}, reentrant phase transition \cite%
{BIRPT} as well as superfluid-like phase transition \cite{superfluidBH} as
well as study of triple points \cite{MRPT}. For more details, we refer to
\cite{BHCh} and references therein.

Criticality is an interesting topic in phase transition context
because thermodynamic quantities of the system show non-analytic
behavior as one approaches the critical point, where a continuous
phase transition occurs. This non-analytic behavior is expressed
in the terms of power law functions which are governed by the
critical exponents. The set of critical exponents define the
universality class of a system, which are thought to obey the same
symmetry principles. In case of charged AdS black hole, a
continuous (second order) phase transition occurs between
small-large black hole in an extended phase space \cite{PV}.
Furthermore, the critical exponents associated with this
transition are the same as those in the Van der Waals liquid-gas
transition, i.e., both systems belong to the same universality
class \cite{PV}. Critical behavior and universality class of AdS
black holes in the extended phase space have been investigated in
various spacetimes \cite{variousBH1,variousBH2}. Recently, authors
of Ref. \cite{Majhi1} have shown that if there is a critical point
in an extended phase space of general AdS black hole, the
small-large black hole transition is in the Van der Waals
universality class. Another interesting approach towards critical
phenomena of AdS black hole is through variation of the electric
charge of black hole in a fixed AdS background geometry, i.e.
keeping the cosmological constant (pressure) as a fixed parameter.
In this perspective, the charge of black hole is treated as a
natural thermodynamic variable that can take on arbitrary values
in the thermodynamic process. In this view, the thermodynamic
behavior of AdS black hole is analyzed in a thermodynamics phase
space. The critical point and associated critical exponents in a
phase space of black hole have been studied in a general way
\cite{Majhi2}. It was found that the values of critical exponents
differ from those of the Van der Waals phase transition
\cite{Majhi2}. {All these phase transitions are also discussed
from the thermogemetrical point of view \cite{Majhi3} where the
phase transition is identified as the divergence of the Ricci
scalar of the thermodynamic geometry}. In additional, an
interesting phenomenon of black hole reentrant phase transition
has also been investigated in four-dimensional Born-Infeld AdS
black in which the charge of system can vary and the cosmological
constant is fixed \cite{AA}.

On the other hand, an alternative view of phase space has been proposed more
recently where various thermodynamic quantities find more natural and
physical meaning \cite{AAA}. In this view, the \emph{square} of the electric
charge of black hole is considered as a thermodynamic variable and $%
\Psi=1/2r_{+} $ is subsequently considered as its conjugate, where
$r_{+}$ is the horizon radius. It was indeed shown that in this
alternative phase space, phase transition and critical behavior of
RN-AdS black hole, in four dimensions, occur in the $Q^{2}$-$\Psi
$ plane, where relevant response function clearly signifies stable
and unstable region \cite{AAA}. Remarkably, in this viewpoint, the
small-large black hole phase transition is quite similar to the
Van der Waals liquid-gas system and belongs to the same
universality class in contrast with previous study of
\cite{Majhi2}. Additionally, it would be interesting to study the
universality class and critical properties for any AdS black hole
in an alternative phase space where the cosmological constant
($\Lambda $) is taken to be constant. Hence, in this paper, we
provide a general framework, i.e. a metric independent way, for
investigating the critical behavior of AdS black hole in the
above-mentioned alternative phase space approach in which a
continuous phase transition happens. {In this analysis, the
exsistance of the phase transition with respect to the alternative
phase space variables is assumed to be there}. Based on the
expansion of the equation of state and Landau thermodynamic
potential in the neighborhood of a critical point in the
alternative phase space, we find that a set value of critical
exponents for generic black hole is analogous to usual Van der
Waals system. Also, we show that the scalar curvature in
thermogeometric picture diverges at the critical point of black
hole in an alternative phase space.

The present Article is structured as follows: In the next section, we obtain
the critical exponents by using the equation of state of black hole in a
general scheme. In section \ref{LandauF}, by considering the Landau
thermodynamic potential, we study universality properties of black hole at
phase transition. In section \ref{GTD}, we investigate thermodynamic
geometry of the system at critical point in alternative phase space. The
last section is devoted to concluding remarks.


\section{Equation of state: General approach \label{GS}}

Here, we intend to prove that the critical exponents of a continuous
(second-order) phase transition are independent of the metric function
(black hole) in an alternative phase space where $Q^{2}$ is treated as a
thermodynamic quantity and its conjugate is $\Psi $ as proposed in \cite{AAA}%
. For this purpose, the first law of black hole thermodynamics, for constant
pressure, is written as%
\begin{equation}
dM=TdS+\Psi dq,  \label{1stlaw}
\end{equation}%
where $q=Q^{2}$. Here $M$, $S$ and $T$ are the total mass, entropy and
Hawking temperature of black hole, respectively. In general, the entropy
only depends on the event horizon, $r_{+}$, i.e. $S=S(r_{+})$. The Gibbs
free energy is a thermodynamic potential that can be calculated by Legendre
transform of the above equation%
\begin{equation}
dG=-SdT+\Psi dq,
\end{equation}%
where $G=G(T,q)$. Thermodynamics of black hole may be described by equation
of state $q=q(\Psi ,T)$ where $q$ depends on $\Psi $ and $T$. As we know,
the critical point in $q-\Psi $ plane is characterized by \cite{AAA}%
\begin{equation}
\frac{\partial q}{\partial \Psi }\Big|_{T_{c}}=0,\quad \quad \frac{\partial
^{2}q}{\partial \Psi ^{2}}\Big|_{T_{c}}=0,  \label{Q1}
\end{equation}%
where the subscript $c$ refers to the critical point and critical quantities
are $T_{c}$, $\Psi _{c}$ and $q_{c}$. The behavior of thermodynamic
functions near the critical point is identified by the critical exponents
which are defined for a Van der Waals system as \cite{Callen}%
\begin{eqnarray*}
&&C\sim \left\vert T-T_{c}\right\vert ^{-\alpha }, \\
&&\left\vert v_{l}-v_{g}\right\vert \sim \left\vert T-T_{c}\right\vert
^{\beta }, \\
&&P-P_{c}\sim \left\vert v_{l}-v_{g}\right\vert ^{\delta }, \\
\chi _{_{T}} &=&-\frac{1}{v}\frac{\partial v}{\partial P}\Big|_{T}\sim
\left\vert T-T_{c}\right\vert ^{-\gamma }.
\end{eqnarray*}%
The first equation defines the exponent for thermal response function $C$
(heat capacity), the second equation characterizes the non-analytic behavior
of order parameter at the critical point, the third defines the critical
isotherm, and the last equation defines the singularity in (mechanical)
response function, $\chi _{_{T}}$ (isothermal compressibility). Here, $v$
and $P$ are specific volume and pressure, respectively, which define the
thermodynamic phase space. Now, in order to find the critical exponents for
a black hole, we expand $q$ around the critical point
\begin{eqnarray}
q(\Psi ,T) &=&a_{00}+a_{01}\left( T-T_{c}\right) +a_{02}\left(
T-T_{c}\right) ^{2}  \notag \\
&&+a_{11}\left( T-T_{c}\right) \left( \Psi -\Psi _{c}\right) +a_{12}\left(
T-T_{c}\right) ^{2}  \notag \\
&&\times \left( \Psi -\Psi _{c}\right) +a_{30}\left( \Psi -\Psi _{c}\right)
^{3}+\ldots ,  \label{expantion}
\end{eqnarray}%
where we have used the convention $a_{ij}\equiv \left( i!j!\right)
^{-1}\partial ^{i+j}q/\partial \Psi ^{i}\partial T^{j}\Big|_{\Psi
_{c},T_{c}} $, and use has been made of Eq.({\ref{Q1}}). For simplicity, we
rewrite thermodynamic variables in dimensionless form as%
\begin{eqnarray}
q &=&q_{c}(1+\varrho ),  \notag \\
T &=&T_{c}(1+t),  \notag \\
\Psi &=&\Psi _{c}(1+\psi ),  \label{redefine}
\end{eqnarray}%
where $\varrho $, $t$ and $\psi $ are deviation from critical point.
Substituting above expressions into Eq.(\ref{expantion}), we have%
\begin{equation}
\varrho =a_{01}^{\prime }t+a_{02}^{\prime }t^{2}+a_{11}^{\prime }t\psi
+a_{12}^{\prime }t^{2}\psi +a_{30}^{\prime }\psi ^{3},  \label{cpdev}
\end{equation}%
in which the prime quantities are equal to the rescaled coefficients in Eq.(%
\ref{expantion}). Since phase transition occurs between small and large
black hole under constant charge, one writes
\begin{eqnarray}
\varrho &=&a_{01}^{\prime }t+a_{02}^{\prime }t^{2}+a_{11}^{\prime }t\psi
_{s}+a_{12}^{\prime }t^{2}\psi _{s}+a_{30}^{\prime }\psi _{s}^{3}  \notag \\
&=&a_{01}^{\prime }t+a_{02}^{\prime }t^{2}+a_{11}^{\prime }t\psi
_{l}+a_{12}^{\prime }t^{2}\psi _{l}+a_{30}^{\prime }\psi _{l}^{3},
\label{Eq1}
\end{eqnarray}%
here, $\psi _{s}$ ($\psi _{l}$) stands for $\psi $ at small (large) horizon.
Also, applying the Maxwell's equal area law, $\oint \Psi dq=0$, which is
obtained from Gibbs free energy \cite{AAA,Callen}, and using Eq.(\ref{cpdev}%
) one can arrive at
\begin{equation}
\int_{\psi _{l}}^{\psi _{s}}\psi \Big(a_{11}^{\prime }t+a_{12}^{\prime
}t^{2}+3a_{30}^{\prime }\psi ^{2}\Big)d\psi =0.  \label{Maxr}
\end{equation}%
It is a matter of calculations to show that Eqs.(\ref{Eq1}) and (\ref{Maxr})
have the following solution
\begin{equation}
\psi _{l}=-\psi _{s}=\sqrt{-\frac{a_{11}^{\prime }t+a_{12}^{\prime }t^{2}}{%
a_{30}}}.
\end{equation}%
Therefore, the order parameter near the critical point behaves as%
\begin{equation}
\left\vert \psi _{s}-\psi _{l}\right\vert =2\psi _{s}\sim
t^{1/2}\Longrightarrow \beta =1/2.
\end{equation}%
In the vicinity of critical point, the shape of the critical isotherm, $t=0$%
, is obtained by%
\begin{equation}
\varrho =a_{30}^{\prime }\psi ^{3}\Longrightarrow \delta =3.
\end{equation}%
According to equation (\ref{cpdev}), the behavior of response function, $%
\chi _{_{T}}=\partial \Psi /\partial q\Big|_{T}$, is given by%
\begin{equation}
\chi _{_{T}}\sim \frac{1}{a_{11}^{\prime }t}\Longrightarrow \gamma =1.
\end{equation}

To denote the specific heat at fixed $\Psi $ close to the critical point, we
perform the expansion of entropy around the critical point%
\begin{eqnarray}
S(\Psi ,T) &=&s_{00}+s_{01}(T-T_{c})+s_{10}\left( \Psi -\Psi _{c}\right)
\notag \\
&&+s_{11}\left( T-T_{c}\right) \left( \Psi -\Psi _{c}\right) +\ldots .
\end{eqnarray}%
where $s_{ij}\equiv \left( i!j!\right) ^{-1}\partial ^{i+j}S/\partial \Psi
^{i}\partial T^{j}\Big|_{\Psi _{c},T_{c}}$. With $S$ at hand, we can extract
the critical exponent $\alpha $, as follows
\begin{equation*}
C_{\Psi }=T\frac{\partial S}{\partial T}\Big|_{\Psi
}=T_{c}s_{01}\Longrightarrow \alpha =0.
\end{equation*}%
It is remarkable to note that $S$ is only a function of $\Psi $ for Maxwell
electrodynamics, i.e. $s_{ij}=0$ for $j\neq 0$ \cite{AAA}. In this way we
calculate the critical exponents of a black hole in a general framework
without specifying the form of the metric, which coincide with those
obtained for Van der Waals fluid system.

\section{Phenomenological aspect: Landau function\label{LandauF}}

Let us define the thermodynamic potential in this case as
\begin{equation}
K=M-TS-\Psi q~.  \label{G}
\end{equation}%
Then use of first law Eq.(\ref{1stlaw}) yields
\begin{equation}
dK=-SdT-qd\Psi ~.  \label{dG}
\end{equation}%
The above states that $K$ is function of both $T$ and $\Psi $ and
\begin{equation}
S=-\Big(\frac{\partial K}{\partial T}\Big)_{\Psi };\,\,\,\ q=-\Big(\frac{%
\partial K}{\partial \Psi }\Big)_{T}~.  \label{Sq}
\end{equation}%
Consequently, the conditions Eq.(\ref{Q1}) at the critical point take the
following forms:%
\begin{equation}
\frac{\partial ^{2}K}{\partial \Psi ^{2}}\Big|_{T_{c}}=0,\quad \quad \frac{%
\partial ^{3}K}{\partial \Psi ^{3}}\Big|_{T_{c}}=0.  \label{G2G3}
\end{equation}

Now since $K=K(T,\Psi )$, Taylor expansion of it near the critical point is
given by
\begin{eqnarray}
&&K(T,\Psi )=b_{00}+b_{10}(T-T_{c})+b_{11}(T-T_{c})(\Psi -\Psi _{c})  \notag
\\
&&+b_{01}(\Psi -\Psi _{c})+b_{20}(T-T_{c})^{2}+b_{21}(T-T_{c})^{2}(\Psi
-\Psi _{c})  \notag \\
&&+b_{22}(T-T_{c})^{2}(\Psi -\Psi _{c})^{2}+b_{12}(T-T_{c})(\Psi -\Psi
_{c})^{2}  \notag \\
&&+b_{13}(T-T_{c})(\Psi -\Psi _{c})^{3}+b_{04}(\Psi -\Psi _{c})^{4}+\dots ~.
\label{GT}
\end{eqnarray}%
In the above we have used $b_{ij}=\left( i!j!\right) ^{-1}(\partial
^{i+j}K)/(\partial ^{i}T\partial ^{j}\Psi )\Big|_{\Psi _{c},T_{c}}$and the
condition Eq.(\ref{G2G3}). Using Eq.(\ref{redefine}), we obtain the near
critical point $K$ as
\begin{eqnarray}
&&K(t,\psi )=b_{00}+b_{10}^{\prime }t+b_{11}^{\prime }t\psi +b_{01}^{\prime
}\psi +b_{20}^{\prime }t^{2}  \notag \\
&&+b_{12}^{\prime }t\psi ^{2}+b_{21}^{\prime }t^{2}\psi +b_{22}^{\prime
}t^{2}\psi ^{2}+b_{13}^{\prime }t\psi ^{3}+b_{04}^{\prime }\psi ^{4}~.
\label{Gfinal}
\end{eqnarray}%
Here the prime quantities are the rescaled coefficients which appear in Eq.(%
\ref{GT}). Since their explicit forms are not needed, we do not mention
their values. The same logic will be followed again and again. Using the
first and second relations of Eq.(\ref{Sq}), we obtain
\begin{eqnarray}
&&S=b_{10}^{\prime \prime }+b_{11}^{\prime \prime }\psi +b_{20}^{\prime
\prime }t+b_{21}^{\prime \prime }t\psi +b_{12}^{\prime \prime }\psi ^{2}
\notag \\
&&+b_{22}^{\prime \prime }t\psi ^{2}+b_{13}^{\prime \prime }\psi ^{3}~,
\label{exentropy} \\
&&q=a_{01}^{\prime \prime }+a_{11}^{\prime \prime }t+a_{12}^{\prime \prime
}t\psi +a_{21}^{\prime \prime }t^{2}+a_{22}^{\prime \prime }t^{2}\psi  \notag
\\
&&+a_{13}^{\prime \prime }t\psi ^{2}+a_{04}^{\prime \prime }\psi ^{3}~,
\label{q}
\end{eqnarray}%
respectively. In the case of Maxwell electrodynamics, the coefficients of $t$
in Eq.(\ref{exentropy}) is zero i.e. $b_{2i}^{\prime \prime }=0$ where $%
i=0,1,3$.

Using the Maxwell's equal area law, one obtains
\begin{equation}
\int_{\psi _{s}}^{\psi _{l}}\psi dq=\int_{\psi _{s}}^{\psi _{l}}\psi
(a_{12}^{\prime \prime }t+a_{22}^{\prime \prime }t^{2}+2a_{13}^{\prime
\prime }t\psi +3a_{04}^{\prime \prime }\psi ^{2})d\psi =0~.  \label{Maxwell}
\end{equation}%
Near the critical point, $q$ is assumed to be the same for $\psi _{s}$ and $%
\psi _{l}$, i.e.:
\begin{eqnarray}
&&q=a_{01}^{\prime \prime }+a_{11}^{\prime \prime }t+a_{12}^{\prime \prime
}t\psi _{s}+a_{21}^{\prime \prime }t^{2}+a_{22}^{\prime \prime }t^{2}\psi
_{s}  \notag \\
&&+a_{13}^{\prime \prime }t\psi _{s}^{2}+a_{04}^{\prime \prime }\psi
_{s}^{3},  \notag \\
&&q=a_{01}^{\prime \prime }+a_{11}^{\prime \prime }t+a_{12}^{\prime \prime
}t\psi _{l}+a_{21}^{\prime \prime }t^{2}+a_{22}^{\prime \prime }t^{2}\psi
_{l}  \notag \\
&&+a_{13}^{\prime \prime }t\psi _{l}^{2}+a_{04}^{\prime \prime }\psi
_{l}^{3}.  \label{qsl}
\end{eqnarray}%
Equating the above two and using Eq.(\ref{Maxwell}), one obtains
\begin{equation}
\psi _{l,s}=\frac{-a_{13}^{\prime \prime }t\pm \sqrt{3t\left[ a_{13}^{\prime
\prime 2}t-3a_{04}^{\prime \prime }\left( a_{12}^{\prime \prime
}+a_{22}^{\prime \prime }t\right) \right] }}{3a_{04}^{\prime \prime }}~,
\label{beta}
\end{equation}%
which leads to
\begin{equation}
\left\vert \psi _{s}-\psi _{l}\right\vert =\frac{2\sqrt{3t\left[
a_{13}^{\prime \prime 2}t-3a_{04}^{\prime \prime }\left( a_{12}^{\prime
\prime }+a_{22}^{\prime \prime }t\right) \right] }}{3a_{04}^{\prime \prime }}%
~,
\end{equation}%
and to the lowest order in $t$%
\begin{equation}
\left\vert \psi _{s}-\psi _{l}\right\vert \sim (T-T_{c})^{1/2}~,
\end{equation}%
which yields $\beta =1/2$. Next, Eq.(\ref{q}) at $T=T_{c}$ reduces to
\begin{equation}
q\sim \psi ^{3}\sim (\Psi -\Psi _{c})^{3}~,
\end{equation}%
and so $\delta =3$.

The isothermal compressibility is defined as $\chi _{T}=(\partial \Psi
/\partial q)_{T}$. Therefore differntiating Eq.(\ref{q}) we find (to the
lowest order in $t$), $\chi _{T}\sim t^{-1}$ which yields $\gamma =1$.
Similarly, the specific heat is given by $C_{\Psi }=T(\partial S/\partial
T)_{\Psi }$ and can be calculated using Eq.(\ref{exentropy})
\begin{equation}
C_{\Psi }=(1+t)\Big(\frac{\partial S}{\partial t}\Big)_{\psi }=(1+t)\left(
b_{20}^{\prime \prime }+b_{21}^{\prime \prime }\psi +b_{22}^{\prime \prime
}\psi ^{2}\right) ~,
\end{equation}%
which to the lowest order is just a constant. Therefore we find the critical
exponent $\alpha =0$.

\section{Thermogeometric description\label{GTD}}

The approach toward phase transition by considering thermodynamic geometry
(thermogeometric) was first introduced by Weinhold \cite{Weinhold} and
Ruppeinner \cite{Ruppeiner1}. In this case the Ricci scalar of the metric
diverges at the critical point or in other words divergence of Ricci scalar
is the signature of the phase transition \cite{Ruppeiner2,DCP}. Both of
these approaches are conformally related to each other by the temperature $T$%
. For recent studies on the Ruppeiner thermodynamic geometry, see \cite%
{Mirza1,Mirza2,1502.00386,1602.03711,1610.06352} and references therein.
This approach has different properties due to the different thermodynamic
potentials, i.e. is not invariant under Legendre transformation \cite%
{Mrugala,Quevedo}. This distinct problem was later remedied by Quevedo \cite%
{Quevedo}, who presented a Legendre invariant set of metrics in the phase
space. In particular, a Legendre invariant metric has been investigated for
black holes in \cite{Alvarez} (see also \cite{Majhi4}). Below, we follow procedure in \cite{Alvarez},
to construct the Legendre invariant metric in an alternative phase space.

The idea is as follows: First construct a thermodynamic phase space $%
\mathcal{T}$ on which the coordinates are $\mathcal{Z}^{A}=(\mathcal{F},%
\mathcal{X}^{a},\mathcal{P}^{a})$ where $\mathcal{F}$ is the thermodynamic
potential and $\mathcal{P}^{a}$ are the conjugate variables of the
thermodynamic variables $\mathcal{X}^{a}$. In this representation the
fundamental one form is given by
\begin{equation}
\theta _{\mathcal{F}}=d\mathcal{F}-\sum_{a,b}\delta _{ab}\mathcal{P}^{a}d%
\mathcal{X}^{b}~,
\end{equation}%
where $\delta _{ab}$ is the Kronecker delta. With the present setup, one can
choose the thermodynamic geometry on $\mathcal{T}$ as
\begin{eqnarray}
&&\mathcal{G}=\Big(d\mathcal{F}-\sum_{a,b}\delta _{ab}\mathcal{P}^{a}d%
\mathcal{X}^{b}\Big)^{2}  \notag \\
&&+\lambda \Big(\sum_{a,b}\xi _{ab}\mathcal{P}^{a}\mathcal{X}^{b}\Big)\Big(%
\sum_{c,d}\eta _{cd}d\mathcal{P}^{c}d\mathcal{X}^{b}\Big)~,  \label{GL}
\end{eqnarray}%
which is invariant under the following set of Legendre transformations:
\begin{eqnarray}
&&\mathcal{F}_{old}=\mathcal{F}_{new}-\delta _{ab}\mathcal{X}_{new}^{a}%
\mathcal{P}_{new}^{b}~,  \notag \\
&&\mathcal{X}_{old}^{a}=-\mathcal{P}_{new}^{a};\,\,\,\,\ \mathcal{P}%
_{old}^{a}=\mathcal{X}_{new}^{a}~.
\end{eqnarray}%
Here $\eta _{ab}=diag(-1,1,1,\dots )$ and $\lambda $ is an arbitrary
Legendre invariant function of $\mathcal{X}^{a}$ while $\xi _{ab}$ is an
arbitrary diagonal constant matrix. The simplest choice is chosen as $%
\lambda =1$ and $\xi _{ab}=diag(1,1,\dots )$. Therefore the general form of
the simplest Legendre invariant metric is
\begin{equation}
\mathcal{G}=\theta _{\mathcal{F}}^{2}+\Big(\sum_{a,b}\xi _{ab}\mathcal{P}^{a}%
\mathcal{X}^{b}\Big)\Big(\sum_{c,d}\eta _{cd}d\mathcal{P}^{c}d\mathcal{X}^{b}%
\Big)~.  \label{gg}
\end{equation}%
We shall work with this particular form.

For the present case, we start with $K$, Eq.(\ref{G}), as a thermodynamic
potential. So according to Eq.(\ref{dG}), the coordinates of the
thermodynamic phase space are $\mathcal{Z}^{A}=(K,\mathcal{X}^{a},\mathcal{P}%
^{a})$ with $\mathcal{X}^{a}=(\Psi ,T)$ and $\mathcal{P}^{a}=(-q,-S)$. So
the metric Eq.(\ref{gg}) takes the form:
\begin{equation}
\mathcal{G}_{1}=\theta _{K}^{2}+(-ST-q\Psi )(-dSdT+dqd\Psi )~,  \label{G_1}
\end{equation}%
with $\theta _{K}=dK+qd\Psi +SdT$. Use of Eqs.(\ref{dG}) and (\ref{Sq})
leads to the following form:
\begin{equation}
\mathcal{G}_{1}=(TK_{T}+\Psi K_{\Psi })(-K_{\Psi \Psi }d\Psi
^{2}+K_{TT}dT^{2})~,  \label{G1}
\end{equation}%
where we use the conventions $X_{Y}=\partial X/\partial Y$ and $%
X_{YY}=\partial ^{2}X/\partial Y^{2}$. The above metric is two dimensional
which has the general form:
\begin{equation}
ds^{2}=-f(x,y)dx^{2}+g(x,y)dy^{2}~.  \label{ds}
\end{equation}%
The Ricci scalar of this metric is given by
\begin{equation}
R=\frac{1}{2f^{2}g^{2}}\Big[f\Big(f_{y}g_{y}-g_{x}^{2}\Big)+g\Big\{%
f_{y}^{2}-f_{x}g_{x}-2f\Big(f_{yy}-g_{xx}\Big)\Big\}\Big]~.  \label{R}
\end{equation}
Here $f=(TK_{T}+\Psi K_{\Psi })K_{\Psi \Psi }$ and $g=(TK_{T}+\Psi K_{\Psi
})K_{TT}$. Therefore the Ricci scalar $R$ diverges when $K_{\Psi \Psi }=0$
provided the numerator is finite. If both vanish, then one needs to be
careful and use L'Hospital's rule to arrive at the same conclusion. This
shows that the first condition (see Eq.(\ref{G2G3})) at the critical point
implies the divergence of Ricci tensor of the metric Eq.(\ref{G1}) at the
critical point, i.e. $K_{\Psi \Psi }=2b_{02}=0$.

For the other condition we consider the original situation where $q$ is
expressed as function of $\Psi $ and $T$. In this case $dq=q_{\Psi }d\Psi
+q_{T}dT$. Therefore the Legendre invariant metric can be chosen as
\begin{equation}
\mathcal{G}_{2}=\theta _{q}^{2}+(q_{\Psi }\Psi +q_{T}dT)(-d\Psi dq_{\Psi
}+dTdq_{T})~,
\end{equation}%
where we have considered the thermodynamic phase space as $\mathcal{Z}%
^{A}=(q,\mathcal{X}^{a},\mathcal{P}^{a})$ with $\mathcal{X}^{a}=(\Psi ,T)$
and $\mathcal{P}^{a}=(q_{\Psi },q_{T})$. Here the fundamental one form is $%
\theta _{q}=dq-q_{\Psi }d\Psi -q_{T}dT$. Now since $q_{\Psi }$ and $q_{T}$
are functions of both $\Psi $ and $T$, proceeding as before, the above
reduces to the following form:
\begin{equation}
\mathcal{G}_{2}=(q_{\Psi }\Psi +q_{T}dT)(-q_{\Psi \Psi }d\Psi
^{2}+q_{TT}dT^{2})~.  \label{G2}
\end{equation}%
This is again of the form Eq.(\ref{ds}) whose Ricci scalar is given by Eq.(%
\ref{R}). Here $f=(q_{\Psi }\Psi +q_{T}dT)q_{\Psi \Psi }$ and $(q_{\Psi
}\Psi +q_{T}dT)q_{TT}$. Therefore the Ricci scalar for the present metric
diverges at the critical point as $q_{\Psi \Psi }$ vanishes at this point,
i.e. $q_{\Psi \Psi }=2a_{20}=0$.


\section{concluding remarks}

\label{con}

Choosing the correct independent thermodynamic variables is a key starting
point in any thermodynamic treatment. The convenient choice of independent
thermodynamic variables can lead to easier solutions, as is clearly
demonstrated by usefulness of Legendre transform in thermodynamics. However,
the ``wrong" set of independent variables could lead to nonphysical results.
In Ref.\cite{AAA}, an alternative thermodynamic phase space was proposed for
RN black holes in AdS space. In this alternative view, the square of black
hole electric charge (instead of the usual charge) was considered to be the
independent thermodynamic variable. It was shown that the thermodynamic
behavior in such an alternative view makes more physical sense and that the
the critical behavior resembled strongly with Van der Waals fluid, belonging
to the same universality class. In the present work, we have approached the
same problem by (i) generalizing to \emph{any} AdS black hole independent of
spacetime metric, and (ii) solving the problem both from the equation of
state as well as thermodynamic potential approach. Both approaches lead
clearly to the set of four critical exponents which are the same as the Van
der Waals fluid system. This provides further evidence for the generality of
Van der Waals universality class for AdS black holes on one hand, as well as
indicating the validity of the alternative phase space proposed in \cite{AAA}%
. Furthermore, we have also provided a thermogeometric approach where Ricci
scaler has been calculated and shown to diverge at the critical point within
the general alternative phase space.


\begin{acknowledgments}
We thank Shiraz University Research Council. The work of AS has been
supported financially by Research Institute for Astronomy and Astrophysics
of Maragha, Iran. The work of BRM is supported by a START-UP RESEARCH GRANT
(No. SG/PHY/P/BRM/01) from Indian Institute of Technology Guwahati, India.
\end{acknowledgments}


\end{document}